\documentclass[aps, twocolumn, groupedaddress, 10pt]{revtex4-1}   
\usepackage{amsmath,amssymb,graphicx,natbib}
\usepackage{epstopdf}

\begin{document}

\title{Inhomogeneous broadening of optical transitions of $\mathrm{^{87}Rb}$ atoms in an optical nanofiber trap}

\author{J. Lee}\altaffiliation{J. Lee and J. A. Grover equally contributed to this work.}
\author{J. A. Grover}\altaffiliation{J. Lee and J. A. Grover equally contributed to this work.}
\author{J. E. Hoffman}
\author{L. A. Orozco}
\author{S. L. Rolston}\email{rolston@umd.edu}
\affiliation{Joint Quantum Institute, Department of Physics, University of Maryland and National Institute of Standards and Technology, College Park, MD 20742, USA}

\date{\today}

\begin{abstract}
We experimentally demonstrate optical trapping of $\mathrm{^{87}Rb}$ atoms using a two-color evanescent field around an optical nanofiber. In our trapping geometry, a blue-detuned traveling wave whose polarization is nearly parallel to the polarization of a red-detuned standing wave produces significant vector light shifts that lead to broadening of the absorption profile of a near-resonant beam at the trapping site. A model that includes scalar, vector, and tensor light shifts of the probe transition $5S_{1/2}$-$5P_{3/2}$ from the trapping beams, weighted by the temperature-dependent position of the atoms in the trap, qualitatively describes the observed asymmetric profile and explains differences with previous experiments that used Cs atoms. The model provides a consistent way to extract the number of atoms in the trap.
\end{abstract}


\maketitle

\section{Introduction}
\label{intro}

The small mode volume of evanescent field atom traps engenders strong atom-light interactions without the need for a cavity~\cite{LeKien05}. The recent demonstrations of trapping $^{133}$Cs with an optical nanofiber (ONF)~\cite{Vetsch10, Beguin14} - and a state-insensitive variant~\cite{Goban12} mark an important experimental realization of these systems. Their high optical depth ($OD$) allows for efficient dispersive readout~\cite{Dawkins11} and strong nonlinear interactions, and they could potentially realize collective effects such as superradiance~\cite{LeKien08}. The success of ONF traps has inspired a growing effort to trap atoms in the evanescent field of nanophotonic waveguides~\cite{Lee13b, Hung13}. This regime of strong coupling opens the door to the study of long-range interactions and the formation of so-called atomic mirrors~\cite{Chang12}, the observation of self-crystallization~\cite{Chang13, Ritsch13}, or the generation of a sub-Poissonian atom number distribution~\cite{Beguin14}. Furthermore, low loss ONFs~\cite{Ward14, Hoffman14, Aoki14} have been proposed as one of steps toward the realization of a hybrid quantum system coupling photons to atoms to superconducting circuit elements~\cite{Verdu09, Hoffman11, Kim11, Hafezi12, Rolston13, Lee13a, Voigt15}.


Optical dipole trapping of atoms is a well-developed technology applied to numerous atomic species.  The extension of optical trapping to evanescent fields of an ONF shares similarities with dipole trapping with free space beams, but has one distinction - the evanescent field may have a substantial longitudinally polarized component of the electric field.  This can lead to surprisingly large differences in the absorption of probe light for two different species, even when they are both alkali atoms (e.g. Rb and Cs), due to the effects of the vector light shift.

Our system traps atoms with two lasers, achieving trap depths of a few hundreds of microKelvin. We cannot simply determine atom number by the absorption of a probe beam by an optically thick medium with a Lorentzian line shape.  Distinct asymmetries are observed that we  trace to the effects of the vector light shifts associated with the optical trapping fields, and their inherent elliptical polarization with an appreciable component along the direction of propagation.  Although Rb and Cs are nominally atoms with very similar atomic structure, the light shifts can in fact be quite different, with differential light shifts much larger in Rb than Cs, leading to a modified  absorption profile.  


The paper is organized as follows. Sec.~\ref{setup} outlines the experimental setup. We present experimental confirmation of our trap in Sec.~\ref{exp}. We introduce a theoretical model based on light shifts, finite atom temperature, and population redistribution in Sec.~\ref{theory}, and use it to study the inhomogeneous absorption profile. Sec.~\ref{conc} summarizes our findings and provides an experimental outlook. 

\section{Setup}
\label{setup}

\begin{figure}
\centering
\includegraphics[width=1\columnwidth]{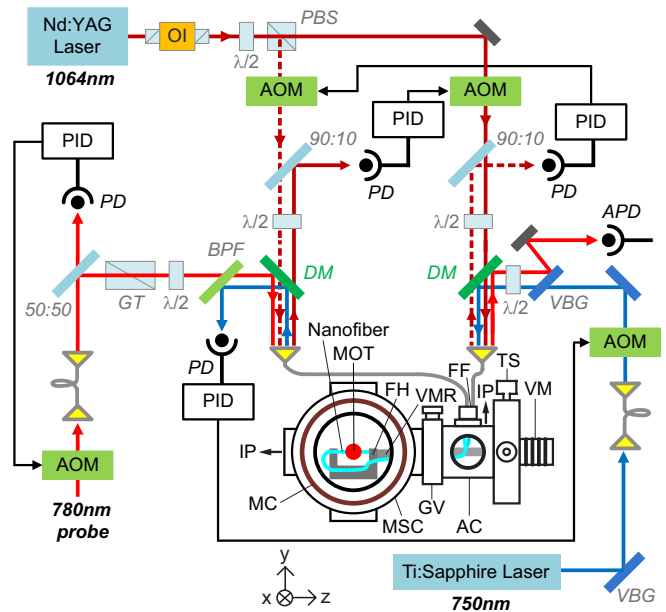}
\caption{Experimental schematic. For $^{87}$Rb atoms, the standing waves with Nd:YAG laser (1064\,nm) and the travelling wave with Ti:Sapphire laser (750\,nm) create a nanofiber atom trap. An absorption signal is measured by a sub-pW 780\,nm probe. See text for details. AC: antechamber; AOM: acousto-optic modulator; APD: avalanche photodiode; BPF: band-pass filter; DM: dichroic mirror; FH: fiber holder; FF: fiber feedthrough; GT: Glan-Thompson polarizer; GV:  gate valve; IP: ion pump; $\lambda/2$: half-wave plate; MC: MOT coil; MOT: magneto-optical trap; MSC: main science chamber; OI: optical isolator; PBS: polarizing beam spliter; PD: photo diode; PID: proportional-integral-derivative lock box; TS: 2-D translation stage; VBG: volume Bragg grating; VM: vacuum manipulator; VMR: vacuum manipulator rod.}
\label{fig_setup}
\end{figure}

Our system consists of the main science chamber (MSC) with the antechamber (AC) equipped with a precision vacuum manipulator (VM) as shown in Fig.~\ref{fig_setup}. We transfer ONFs from the antechamber to the science chamber without breaking the science chamber vacuum. The science chamber maintains a pressure of $10^{-9}$\,mbar with two ion pumps ($50\,\mathrm{L \cdot s^{-1}}$). We produce the ONF by thinning a single-mode fiber (Fibercore) in an in-lab setup~\cite{Hoffman14}. The ONFs typically have greater than 99\,\% transmission for waists with diameters of 500\,nm and lengths from 1\,mm to 10\,cm. The data presented in this paper are taken using an ONF with a waist of ~530\,nm diameter and 7\,mm length, with a tapering region of 39\,mm in length. The produced ONF epoxied onto the titanium alloy fiber holder (FH) is held on the vacuum manipulator rod (VMR) at the antechamber and is transfered to the science chamber. The mounted ONF is extended to the out-of-vacuum patch-cord single-mode fiber via the Teflon ferrule fiber feedthrough (FF) with a Swagelok.

A magneto-optical trap (MOT) loaded from a background vapor of $\mathrm{^{87}Rb}$ produces a cloud of $\sim 10^8$ atoms. We overlap the cloud with the ONF waist using magnetic field shim coils and the vacuum manipulator (VG Scienta Transax) with 2-D manual translation stages (TS). Two orthogonal imaging systems ensure alignment. Atoms are loaded into the ONF trap (left on throughout the experiment)  after 90\,ms of increased MOT detuning and a 1\,ms duration optical molasses stage. The sub-Doppler cooling during this loading stage yields MOT temperatures of $\mathrm{\sim15}\,\mu K$, as determined by time-of-flight measurements.

An ONF trap requires light tuned red of resonance (with respect to  to the $^{87}$Rb D2 line) to provide an attractive potential and light tuned blue of resonance to prevent atoms from  striking the ONF surface.  A 750-nm wavelength laser (Coherent Ti:Sa 899) provides the repulsive force, and a 1064-nm wavelength beam (JDSU NPRO) in a standing wave configuration (to provide longitudinal confinement) provides the attractive potential. A potential minimum of 370\,$\mu K$ in depth is formed $\sim$215\,nm from the fiber surface, and the trapping frequencies are calculated to be ($\nu_{r}$, $\nu_{z}$, $\nu_{\phi}$) = (253, 371, 104)\,kHz. All of the ONF trapping beams and the near-resonant probe beam are intensity-stabilized. The amplitude lock of the probe beam includes a sample-and-hold to stabilize the power in between pulses so that the servo can rapidly recapture when the pulse is turned on, minimizing transients.

We measure atomic absorption  with a weak, near-resonance beam (780\,nm) coupled through the ONF, counting transmitted photons with avalanche photodiodes (APD, Laser Components COUNT-250C-FC) operating in Geiger mode. Because light levels near 10\,pW saturate the APDs, great care must be taken to filter stray light and maintain low probe power. Three narrow-line volume Bragg gratings (VBG, OptiGrate BP-785, 0.18\,nm spectral bandwidth at 785\,nm) filter amplified spontaneous emission from the Ti:Sapphire laser near 780\,nm. A VBG at the output of the nanofiber serves as a mirror to direct signal to the APDs and as another filter to block in-fiber background induced by the blue trapping beam. This light due to either fluorescence or Raman scattering is the main source of background in the experiment. Two more bandpass filters further reduce background counts, and finally long-pass color filters (Thorlabs, FGL645) directly in front of the APD fiber couplers reduce short-wavelength background from stray light. A series of differing optical depth neutral density filters before and after the nanofiber allow us to vary the probe intensity while keeping light levels within the dynamic range  of the APDs. TTL pulses from the APDs are counted with a field-programmable gate array (FPGA) and processed to extract absorption signals and full photon counting statistics.

\section{Experiment}
\label{exp}

\begin{figure}
\centering
\includegraphics[width=1\columnwidth]{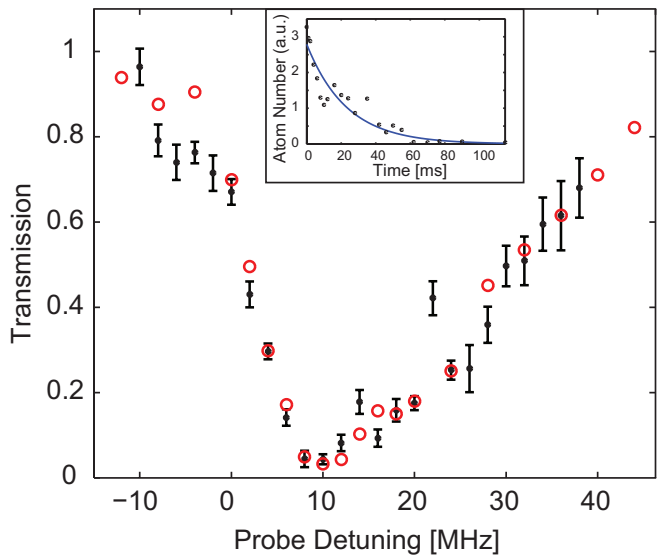}
\caption{Atomic absorption signals as a function of probe detuning relative to the bare atomic transition (black-dot: counted photon numbers with error bars of 1$\sigma$ statistical error; red-circle: averaged absorption signal with 50\,shots). Inset illustrates a typical trap lifetime measurement with a fit to a decaying exponential ($\tau$ = 23\,ms). The absorption measurement is done with red-detuned (3.35\,mW\,$\times$2, standing waves) and blue-detuned (7.4\,mW, travelling wave) trapping beams. The trap configuration is close to the PP case with an angle (23.5$^{\circ}$).}
\label{fig_absorption}
\end{figure}

The absorption profiles are measured via an in-fiber analog of standard absorption spectroscopy. We use two probe pulses; the first pulse measures the atomic absorption signal ($P_{at}$), and the second pulse is a reference signal with no atoms ($P_{0}$). In between the two probe pulses, the 1064\,nm trapping beam is turned off and a slightly blue-detuned laser from the MOT beam paths kicks away the trapped atoms. Based on the ratio of these probe signals, we calculate the measured transmission $T = (P_{at} - P_{bg})/(P_{0} - P_{bg})$, where $P_{bg}$ is the background APD signal with no probe light, with contributions from   detector dark counts and fiber-induced fluorescence. For a single Lorentzian lineshape with width $\Gamma$, one can easily estimate the optical depth $OD$ by fitting $T(\omega) = \exp[ -  OD / ( 1 + 4(\omega-\omega_0)^2/\Gamma^2 ) ]$ to the data. The total number $N$ of trapped atoms is then given by $OD/OD_{1}$, where $OD_{1}$ is the single-atom optical depth. We calculate $OD_1$ to be $2.78\%$ by comparing the atomic absorption cross section to the optical nanofiber mode area. Our measured absorption profile displays a markedly asymmetric lineshape (see Fig.~\ref{fig_absorption}). Therefore, it is not trivial to estimate the number of trapped atoms. For a given number of trapped atoms, any broadening mechanisms serve to reduce the maximum absorption. In addition, the maximum absorption point sets a lower limit on the $OD$ and the number of trapped atoms, i.e. $OD_{low} = -{\rm Ln}[T]$. We will discuss this more in Sec. \ref{theory}, where we develop a method to estimate the number of trapped atoms based on these asymmetric absorption profiles. 

We observe trapping lifetimes (without any additional cooling) of approximately 23\,ms (see Fig. \ref{fig_absorption} (Inset)). 
We observe that our ONF exhibits large-amplitude transverse vibrations near 550\,Hz. These are too low in frequency to produce heating as it would be adiabatic motion in terms of the optical trapping potential. Though this is far away from any relevant trap frequency, the acceleration of the fiber may be high enough at times to affect loading, as the macroscopic motion of the trap is no longer adiabatic relative to the mean atomic motion. This lifetime is typical of ONF traps, which are generally shorter-lived than standard optical dipole traps, for reasons that are not yet fully understood.

Different polarization configurations of the red- and blue-detuned trapping beams can provide trapping, including parallel polarization (PP) and cross polarization (CP). The PP configuration requires less blue-detuned light than the CP configuration, but has larger vector light shifts. The CP configuration results in more azimuthally localized potentials and smaller vector light shifts. In this paper, we chose a trapping geometry close to the PP configuration with an angle (23.5$^{\circ}$) between the polarizations of red- and blue-detuned trapping beams where we measured the deepest optical depth ($\sim$98\% absorption), better  than with the PP or CP configurations (50$\sim$70\% absorption). We analyzed the polarizations of nanofiber modes with Rayleigh scattering, confirming the local polarization angles, but do not have a reason why this particular polarization angle produced the best optical depth.

\section{Model and simulation}
\label{theory}

\begin{figure}
\centering
\includegraphics[width=1\columnwidth]{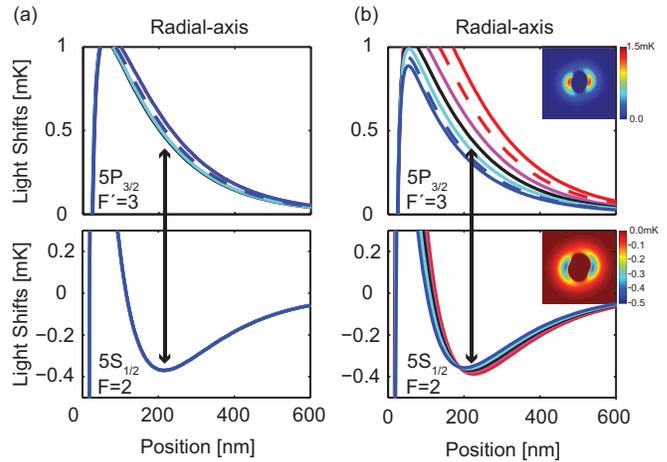}
\caption{Light shifts (radial-axis) with red-detuned (3.35\,mW\,$\times$2) and blue-detuned (7.4\,mW) trapping beams. The trap configuration is close to the PP case with an angle (23.5$^{\circ}$). (a) Light shifts with linearly polarized trapping beams of $\mathrm{5P_{3/2}}$'s $|\mathrm{F'}=3,\,\mathrm{m_{F'}}\rangle$ state and $\mathrm{5S_{1/2}}$'s $|\mathrm{F}=2,\,\mathrm{m_{F}}\rangle$ state. Zeeman sub-levels of $\mathrm{5S_{1/2}}$ are degenerate because of no vector and tensor light shifts, and Zeeman sub-levels of $\mathrm{5P_{3/2}}$ are split due to tensor light shifts. (b) Light shifts with circularly polarized trapping beams of $\mathrm{5P_{3/2}}$'s $|\mathrm{F'}=3,\,\mathrm{m_{F'}}\rangle$ state and $\mathrm{5S_{1/2}}$'s $|\mathrm{F}=2,\,\mathrm{m_{F}}\rangle$ state. Inset shows the cross-sectional view of light shifts. In Fig.~\ref{fig_light_shift}, the color lines of red, red-dash, purple, black, cyan, blue-dash, and blue correspond to $\mathrm{5P_{3/2}}$'s $|\mathrm{F'}=3,\,\mathrm{m_{F'}} = +3, +2, +1, 0, -1, -2, -3 \rangle$ state, respectively, and the lines of red, purple, black, cyan, and blue correspond to $\mathrm{5S_{1/2}}$'s $|\mathrm{F}=2,\,\mathrm{m_{F}} = +2, +1, 0, -1, -2 \rangle$ state, respectively. In addition, the up-down arrow means the light-shifted optical transition frequency at the trapping position. Here, the quantization axis is defined along the propagation direction ($\mathbf{z}$) of the probe beam (see Fig.~\ref{fig_setup}), and the light polarization is defined in the $\mathbf{xy}$ plane with Stokes vectors; the linearly (circularly) polarized light means $S_{3}/S_{0}$ = 0 ($\pm 1$).}
\label{fig_light_shift}
\end{figure}

To calculate the inhomogeneous broadening of the absorption line, we need to include the light-shifted optical transition frequency for atoms trapped in the optical fiber potential. This requires appropriate weighing over the polarization of the modes, the $m$-state distribution of the atoms, and the location of the atoms within the trap due to thermal motion. Qualitatively, lower temperature atoms moves the absorption profile to the blue as the atoms spend more time in a deeper trap, which leads to more light shifts, and higher temperature atoms determine the left-wing of the profile by the Boltzmann truncation. The light-shifted optical transition frequencies of $\mathrm{^{87}Rb}$ atoms at trapping positions move the absorption profile to the blue, and the vector light shifts broaden the blue-side of the profile. 

\begin{figure}
\centering
\includegraphics[width=1\columnwidth]{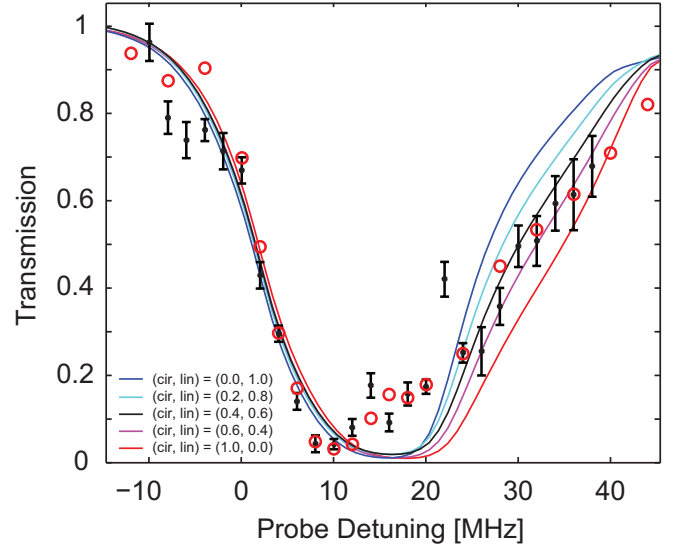}
\caption{The atom number estimation profiles (red-, magenta-, black-, cyan-, blue-line) for asymmetric absorption measurement data (black-dot, red-circle) are represented for trapped atoms with the ratio, (cir, lin), of the circularly polarized light ($\mathrm{S_{3}/S_{0}} = \pm1$) and the linearly polarized light ($\mathrm{S_{3}/S_{0}} = 0$); the quantization axis is defined along the propagation direction of the probe beam. For the fittings, we assume red-detuned (3.35\,mW\,$\times$2) and blue-detuned (7.4\,mW) trapping beams and the trap configuration close to the PP case with an angle (23.5$^{\circ}$).}
\label{fig_fitting}
\end{figure}

The vector light shift is related to the ellipticity of the light and the populated atomic states. When the light propagation direction and the populated atomic state $|\mathrm{F},\,\mathrm{m_{F}}\rangle$ are collinear along the quantization axis ($\mathbf{z}$), the ellipticity is defined in the $\mathbf{xy}$ plane with Stokes vector components $S_k$~\cite{Jackson98}, and the vector light shift ($\hat{H}_{1} \propto C^{(1)}S_{3}F_{z}$) is proportional to the ellipticity of the light, where $C^{(1)}$ is calculated from Wigner-Eckart theorem. The linearly polarized light ($S_{3}=0$) can be symmetrically decomposed into right-hand and left-hand circularly polarized lights ($S_{3} = +1, -1$, respectively) that cancel the ellipticity. The $\mathrm{HE_{11}}$ fundamental mode has an axial-direction ($\mathbf{z}$) electric field component that leads to ellipticity ($S_{3} \neq 0$) in the $\mathbf{xy}$ plane, which creates the vector light shifts (see Fig.~\ref{fig_light_shift} (b)). Instead of considering all the ellipticities for each trapping location, we add linearly and circularly polarized light shifts with a variable ratio (see Fig.~\ref{fig_fitting}). 


A complete description of the atomic absorption has contributions from homogeneous (natural linewidth) broadening $\mathcal{L}_{0}(\omega-\omega')$ and inhomogeneous broadening $n(\omega')$, generally resulting in the symmetric Voight profile~\cite{Demtroder03} as follows:
\small
\begin{eqnarray}
I(\omega) &=& I_0 \int n(\omega') \mathcal{L}_{0}(\omega-\omega') d\omega',
\label{eqn1}
\end{eqnarray}
where $I(\omega)$ is the convolution of Lorentzian and Gaussian profiles.
 For the optical transition of $^{87}$Rb atoms, we study the inhomogeneous broadening to consider the effects of  the atomic temperature distribution, Zeeman-sublevel-dependent population distribution, and the light shifts of the hyperfine ground state ($\mathrm{5S_{1/2}}$ to $\mathrm{5P_{1/2}}$ and $\mathrm{5P_{3/2}}$) and the optical excited state ($\mathrm{5P_{3/2}}$ to $\mathrm{(4\textendash6)D_{3/2}}$, $\mathrm{(4\textendash6)D_{5/2}}$, $\mathrm{(5\textendash8)S_{1/2}}$
)~\cite{Safronova07}. This requires considering scalar, vector, and tensor light shifts~\cite{Mabuchi06, Deutsch10, Kien13}. The vector light shift can be large due to a non-negligible axial-direction electric field component ($E_{z}$) in the fundamental mode $\mathrm{HE_{11}}$~\cite{Kien13}. Figure~\ref{fig_light_shift} shows the light shifts of the hyperfine ground state and the optical excited state, considering the polarization of the blue- and red-detuned trapping beams. The up-down arrow represents the light-shifted optical transition frequency at the trapping position, which affects the absorption profile. The linearly polarized trapping beams have no vector light shifts of $\mathrm{5S_{1/2}}$ and $\mathrm{5P_{3/2}}$ (Fig.~\ref{fig_light_shift} (a)). However, the nanofiber's $\mathrm{HE_{11}}$ mode always has the propagation direction components related to the vector light shift (Fig.~\ref{fig_light_shift} (b)), and moving atoms in the trap with a finite temperature experiences the variation of light shifts. In the experiment, the quantization axis is not well-defined with a residual magnetic field after optical molasses cooling, and the initial states of trapped atoms are assumed to be unpolarized. The PP trap configuration with an angle induces more vector light shifts from the blue-detuned traveling wave than the red-detuned standing waves. In the case of the absorption profile of the unpolarized atoms requiring the sum of all the Zeeman sub-levels' absorptions, $\mathrm{^{87}Rb}$ atoms are expected to experience more asymmetric profile than $\mathrm{^{133}Cs}$ atoms due to the vector light shifts from the upper state manifolds. Because of differences in the atomic structure of higher lying states in Rb compared to Cs, there is no available magic wavelength and excited state vector light shifts are considerably larger. 

For a ground state $|\mathrm{n,\,F,\,m_{F}}\rangle$ and an excited state $|\mathrm{n',\,F',\, m_{F'}}\rangle$ represented by $i$ and $j$, the inhomogeneous term $n_{ij}(\omega)$ can be defined for trapped atoms having a  temperature $T$ as follows:
\small
\begin{eqnarray}
n_{ij}(\omega) = \int_{V_{eff}} \frac{1}{Z} \exp\left(-\frac{U_{ij}(\vec{r})}{k_B T}\right) \delta(\omega - \omega_{ij}(\vec{r})) dV
\label{eqn2}
\end{eqnarray}
where $Z = \int_{V_{eff}} \exp\left(-U_{ij}(\vec{r})/(k_B T)\right) dV$; $U_{ij}(\vec{r})$ is the trapping potential of hyperfine ground states ($\mathrm{5S_{1/2}}$); and $\omega_{ij}(\vec{r})$ is the light-shifted optical transition frequency ($\mathrm{5S_{1/2}}$ to $\mathrm{5P_{3/2}}$ transition). $U_{ij}(\vec{r})$ and $\omega_{ij}(\vec{r})$, dependent on powers and polarizations of the two  trapping beams, have spatial dependence and need to be integrated over the effective volume of a trap site. The atoms with a  temperature $T$ higher than a local trap potential $|U_{ij}(\vec{r})|/k_{B}$ are truncated in the calculation.

Here, we define a homogeneous profile including light shift broadening as follows:
\begin{eqnarray}
\mathcal{L}(\omega-\omega') = \frac{1}{1 + (\omega - \omega')^2 / (\Gamma/2 + \Delta\Gamma(\omega')/2)^2},
\label{eqn3}
\end{eqnarray}
where $\Delta\Gamma(\omega')/2  = \Delta \omega'$ is the broadened width of an optical transition $\omega'$; the standard deviations of state-dependent light-shifted optical transitions at each location $\vec{r}$ are calculated for a frequency $\omega'$ and averaged over all $V_{eff}$.
We define transmission $T(\omega)$ (Sec. \ref{exp}) based on our definition of the inhomogeneous broadening of the optical transition as follows: 
\small
\begin{eqnarray}
T(\omega) &=& \exp[-OD \sum_{i,j} |\tilde d_{ij,\,q}|^2 f_{i} \int n_{ij}(\omega') \mathcal{L}(\omega-\omega') d\omega'],\\
                 &\approx& \exp[- N \cdot OD_{1} \int n_{ij}(\omega') \mathcal{L}(\omega-\omega') d\omega'],
\end{eqnarray}
where $n_{ij}(\omega')$ and $\mathcal{L}(\omega-\omega')$ are defined in Eqn. (\ref{eqn2}, \ref{eqn3}), and $f_{i}$ is the population of Zeeman sub-levels determined by optical Bloch equations during optical pumping from the probe beam. $|\tilde d_{ij,\,q}|^2$ is the relative strength of the atomic dipole moment related to the polarization state $q$ of the probe, and the OD per atom is $OD_{1} = \sigma_0 / A_{\mathrm{eff}}$ ($A_{\mathrm{eff}} = P_{\mathrm{prob}}/I_{\mathrm{prob}}(\vec{r}) = 4.88\,\mathrm{\mu m}^2$: the effective mode area of the nanofiber probe, $\sigma_{0} = 0.1356\,\mathrm{\mu m}^2$: atomic scattering cross-section). This can be regarded as a constant for a given $i, j, q$, and assumes no light shifts from the low intensity probe.

Given the uncertainties in the exact polarization profile of the optical modes where the atoms are trapped, the $m$-state distribution of the atoms, and the degree to which a truncated Boltzmann distribution is a correct assumption, we qualitatively use the asymmetric profiles to  estimate trapped atom number (see Fig. \ref{fig_fitting}). Lower $T$ moves the profile to the blue as the atoms stay more in higher intensity regions. A larger fraction of circular polarization broadens the blue-side of the profile by the vector light shifts. Our calculated profiles do not reproduce the feature of less absorption around 15\,MHz. This may be related to probe-induced heating near resonance or probe-induced optical pumping that modifies the sublevel populations and thus absorption cross section. We focus on the left and right tails of the absorption profile to estimate atom number $N = 302$ for $T$ = 55\,$\mu$K; this corresponds to $OD$ = 8.4 with our calculated $OD_{1}$ = 0.0278. Based on our highest absorption of 96.8\,\% ($OD$ = 3.44) at a probe-detuning of 10\,MHz, the absolute lower bound of trapped atom number is $N$ = 123 (as all broadening mechanisms will decrease the maximum observed absorption and increase the extracted number of atoms).

\section{Conclusion}
\label{conc}

We realized an optical nanofiber atom trap for $^{87}$Rb using the evanescent fields of 750\,nm and 1064\,nm beams, and find absorption profiles with much less symmetry than have been observed with Cs atoms. We qualitatively explain the asymmetric broadening behavior as arising from  larger differential light shifts arising from the $\mathrm{5P_{3/2}}$ to upper transitions, which are larger in Rb than Cs. Using a model of the broadening, we estimate the number of trapped atoms.\\

\begin{acknowledgements}
This work was funded by NSF PFC@JQI and ARO Atomtronics MURI.
\end{acknowledgements}

\end{document}